\documentclass[amsmath,amssymb,superscriptaddress,twocolumn,aps,prx,longbibliography]{revtex4-2}

\usepackage{empheq}
\usepackage{relsize}
\usepackage{physics}
\usepackage{graphicx}
\usepackage{xcolor}
\usepackage{bm}

\newcommand{\vect}[1]{\boldsymbol{#1}}

\newcommand{\adj}[1]{\mathrm{adj} ({#1})}

\begin{document}

\title{Giant Photon Superbunching from Weak Nonlinearity}

\author{You~Wang}
\thanks{These authors contributed equally to this work.}
\affiliation{Division of Physics and Applied Physics, School of Physical and Mathematical Sciences, Nanyang Technological University, Singapore 637371, Singapore}

\author{Xu~Zheng}
\thanks{These authors contributed equally to this work.}
\affiliation{Division of Physics and Applied Physics, School of Physical and Mathematical Sciences, Nanyang Technological University, Singapore 637371, Singapore}

\author{Timothy C.~H.~Liew}
\email{timothyliew@ntu.edu.sg}
\affiliation{Division of Physics and Applied Physics, School of Physical and Mathematical Sciences, Nanyang Technological University, Singapore 637371, Singapore}
\affiliation{MajuLab, International Joint Research Unit UMI 3654, CNRS, Universit\'e C\^ote d'Azur, Sorbonne Universit\'e, National University of Singapore, Nanyang Technological University, Singapore}

\author{Y.~D.~Chong}
\email{yidong@ntu.edu.sg}
\affiliation{Division of Physics and Applied Physics, School of Physical and Mathematical Sciences, Nanyang Technological University, Singapore 637371, Singapore}
\affiliation{Centre for Disruptive Photonic Technologies, School of Physical and Mathematical Sciences, Nanyang Technological University, Singapore 637371, Singapore}

\begin{abstract}
  Photon superbunching, which occurs when the second-order correlation satisfies $g^{(2)}> 2$, is typically associated with strong optical nonlinearities or collective multi-photon emission processes.  We predict that extreme superbunching can also arise in systems of weakly-nonlinear photonic cavities, via the creation of a squeezed vacuum through interference engineering by fine-tuning inter-cavity couplings and drive parameters.  We present numerical calculations indicating that a system of four photonic resonators containing representative Kerr media can achieve $g^{(2)}(0) = 135$ with a $80\,\text{kHz}$ emission rate. Unlike earlier superbunching schemes, this mechanism is highly compatible with integrated photonic platforms constructed using conventional optical media.
\end{abstract}

\maketitle

\section{Introduction}

Photon correlations are central to quantum optics and underpin a wide range of quantum technologies. They are commonly characterized by the second-order correlation function $g^{(2)}(\tau)$ introduced by Glauber~\cite{Glauber63}: $g^{(2)}(0)=1$ for coherent light, $g^{(2)}(0)=2$ for thermal or chaotic light (which exhibits greater bunching than coherent light) \cite{HBT56}, and $g^{(2)}(0)<1$ for antibunched (e.g., photon-blockaded) light.  Classical setups, such as cascaded rotating ground glasses, can convert laser light into pseudothermal light with altered intensity fluctuation distributions achieving $g^{(2)}(0) \sim 3$~\cite{Zhou17}.  The superbunching regime, $g^{(2)}(0)\gg 2$, is of special interest: photons with such extreme correlations could be useful in applications such as ghost imaging, nonlinear optics, and electron sources \cite{Rasputnyi24, Heimerl24, Lemieux25}.

Superbunching can occur through various quantum phenomena, including spontaneous parametric down-conversion \cite{Elvira2011, Iskhakov12, Walls83, Leon19, Manceau2019}, collective effects like superradiance \cite{Bhatti15, Gulfam18, Jahnke16, Fiedler23}, and intra-emitter quantum effects \cite{Heindel2017, Wang21opt}.  These have been shown to achieve $g^{(2)}(0) \sim 10^2$, though recently a specially-prepared nonlinear photonic crystal fiber has been claimed to reach $g^{(2)}(0) \sim 10^4$~\cite{Qin24}.  Broadly speaking, known methods rely on either strong optical nonlinearities to generate a bright squeezed vacuum \cite{Iskhakov12}, or the presence of quantum matter (e.g., collectively coherent emitters or quantum dots with cascaded emission pathways).  This poses a challenge for using integrated quantum photonic devices to generate superbunched light, because such devices have small form factors and, preferably, utilize conventional optical media featuring only weak intrinsic nonlinearities.

However, superbunching by vacuum squeezing may not strictly require strong nonlinearity.  An analogy may be drawn to the phenomenon of unconventional photon blockade (UPB) \cite{Liew10, Ferretti10, Bamba11, Bamba11_2, Flayac13, Xu14, Lemonde14, Wang21, Wang25}, whereby photon blockade ($g^{(2)}(0) \rightarrow 0$) is achieved in a system of coupled optical modes by carefully designing the mode excitation pathways and how they interfere with one another.  UPB has been demonstrated experimentally \cite{Snijders18, Vaneph18}, and there have been recent theoretical proposals aimed at drastically reducing the nonlinearity strengths required \cite{Wang21}, as well as increasing the antibunching lifetime (i.e., the range of $\tau$ over which $g^{(2)}(\tau)$ is suppressed) \cite{Wang25}.  Could this interference-engineering approach be applied to the opposite purpose---superbunching as opposed to antibunching?  If so, what is the largest $g^{(2)}(0)$ that can be achieved, and what sort of photonic system is required?

In this paper, we predict that a system of coupled photonic modes with realistically weak Kerr nonlinearities can exhibit giant superbunching with $g^{(2)}(0) > 100$, comparable to previous superbunching results based on strong nonlinearities, and possibly even higher.  This is accomplished by engineering destructive interference among coherent driving paths to cancel the classical amplitude $\langle \hat{a}_i \rangle$ in a target mode $i$, leaving the photon emission governed entirely by the quantum fluctuations $\hat{d}_i \equiv \hat{a}_i-\langle \hat{a}_i \rangle=\hat{a}_i$.  Consequently, the pair correlation $\langle \hat{d}_i\, \hat{d}_i \rangle$ dominates over the normal occupation $\langle \hat{d}_i^\dagger \hat{d}_i \rangle$, giving rise to
\begin{equation}
  g_{ii}^{(2)}(0) - 2 \;\propto\; \frac{1}{\langle \hat{a}_i^\dagger \hat{a}_i\rangle} \gg 1.
  \label{eq:scaling_result}
\end{equation}
We develop an analytical theory for this mechanism, and verify it with numerical master equation simulations on a minimal four-mode model.  This superbunching scheme is highly compatible with scalable integrated photonic platforms such as silicon carbide microring resonators~\cite{Yi22,Xing19}, as it does not rely on giant optical nonlinearities or collective emission, nor exotic (e.g., non-passive) couplings between the participating optical modes.  Moreover, the required interferometric conditions can be achieved via standard active control parameters in photonic systems, such as laser frequency detuning and inter-cavity coupling strength.  Such interference-aided superbunching may thus supply a workable design principle for compact quantum light sources.

\section{Coupled-mode model}

We begin by presenting our theoretical framework for studying superbunching.  Consider a system of $N$ coupled photonic modes, each of which may be subject to on-site Kerr nonlinearity, loss, and coherent driving. In the frame co-rotating with the driving frequency, the Hamiltonian is (setting $\hbar = 1$)
\begin{align}
  \mathcal{H} = \sum_{j} &\left[-\Delta_j \hat{a}_j^\dagger \hat{a}_j + U_j \hat{a}_j^\dagger \hat{a}_j^\dagger \hat{a}_j \hat{a}_j + F_j \hat{a}_j^\dagger + F_j^* \hat{a}_j\right] \nonumber \\
  &+ \sum_{j \neq k} J_{jk} \hat{a}_j^\dagger \hat{a}_k,
  \label{eq:H}
\end{align}
where, for each mode $j$, $\Delta_j$ is the frequency detuning (relative to the drive), $U_j$ is the Kerr coefficient, $F_j$ is the drive amplitude, $J_{jk} = J_{kj} \in \mathbb{R}$ is the inter-mode coupling (in the absence of magneto-optic effects), and $\hat{a}_j, \hat{a}_j^\dagger$ are the photon annihilation and creation operators.  We also assign each mode a loss rate $\gamma_j$, which enters via the Lindblad master equation
\begin{align}
  \frac{d\hat{\rho}}{dt} &= -i[\hat{\mathcal{H}}, \hat{\rho}] + \sum_j \gamma_j \mathcal{D}[\hat{a}_j]\hat{\rho},
  \label{eq:S_lindblad} \\
  \mathcal{D}[\hat{o}]\hat{\rho} &\equiv \hat{o}\hat{\rho} \hat{o}^\dagger - \{\hat{o}^\dagger \hat{o}, \hat{\rho}\}/2. \label{eq:disspator}
\end{align}

The steady-state second-order correlation between sites $i$ and $j$, with time delay $\tau\ge0$, is
\begin{equation}
  g^{(2)}_{ij}(\tau)=\lim\limits_{t\rightarrow + \infty} \frac{\langle\hat{a}_j^\dagger(t)\hat{a}_i^\dagger(t+\tau)\hat{a}_i(t+\tau)\hat{a}_j(t)\rangle}{\langle\hat{a}_j^\dagger(t)\hat{a}_j(t)\rangle\langle\hat{a}_i^\dagger(t+\tau)\hat{a}_i(t+\tau)\rangle},
  \label{g2def}
\end{equation} 
with the expectation values taken in the steady state. We will mainly be interested in the on-site correlation $g^{(2)}_{ii}(0)$ and the mean photon number $\langle n_i \rangle = \langle \hat{a}_i^\dagger \hat{a}_i \rangle$.

To separate out the classical response, let
\begin{equation}
  \hat{a}_j = \alpha_j + \hat{d}_j,
  \label{eq:displace}
\end{equation}
where $\alpha_j \in \mathbb{C}$ is the classical steady-state amplitude and $\hat{d}_j$ is a fluctuation operator satisfying $\langle \hat{d}_j \rangle = 0$.  If we take the Heisenberg equation $i\dot{\hat{a}}_j = [\hat{a}_j, \mathcal{H}] - i(\gamma_j/2)\hat{a}_j+\hat{\xi}_j$, where $\hat{\xi}_j$ is the quantum noise operator, applying the steady-state condition $\dot{\alpha}_j = 0$ yields
\begin{equation}
  \left(-\Delta_j - i\frac{\gamma_j}{2}\right) \alpha_j + 2U_j |\alpha_j|^2 \alpha_j + \sum_k J_{jk} \alpha_k + F_j = 0.
  \label{eq:classical}
\end{equation}
This is a set of $N$ coupled cubic equations whose solutions determine the classical mean field.

As shown in Appendix~\ref{sec:supp_Heff}, after extracting this classical contribution, the remaining terms form an effective Hamiltonian
\begin{align}
  \hat{\mathcal{H}}_\mathrm{eff} = \sum_j \bigg[&\tilde{\omega}_j \hat{d}_j^\dagger \hat{d}_j + U_j\!\left(\alpha_j^2 (\hat{d}_j^\dagger)^2 + (\alpha_j^*)^2 \hat{d}_j^2\right) \nonumber \\
  &+ 2U_j\!\left(\alpha_j^* \hat{d}_j^\dagger \hat{d}_j^2 + \alpha_j (\hat{d}_j^\dagger)^2 \hat{d}_j\right) \nonumber \\
  &+ U_j\, \hat{d}_j^\dagger \hat{d}_j^\dagger \hat{d}_j \hat{d}_j\bigg] + \sum_{j \neq k} J_{jk}\, \hat{d}_j^\dagger \hat{d}_k,
  \label{eq:Heff}
\end{align}
where
\begin{equation}
  \tilde{\omega}_j = -\Delta_j + 4U_j|\alpha_j|^2
  \label{eq:omegatilde}
\end{equation}
is a renormalized on-site frequency detuning.  This effective Hamiltonian enters via the master equation
\begin{equation}
  \frac{d\hat{\rho}}{dt} = -i[\hat{\mathcal{H}}_\mathrm{eff}, \hat{\rho}] + \sum_j \gamma_j \mathcal{D}[\hat{d}_j]\hat{\rho}.
  \label{eq:S_master}
\end{equation}
The Lindblad dissipator has the same form as in Eq.~\eqref{eq:disspator}, since the displacement \eqref{eq:displace} is a unitary operation on the bosonic Hilbert space.

To solve this quantum Langevin equation analytically, we can linearize $\hat{\mathcal{H}}_\mathrm{eff}$ by retaining only terms quadratic in $\hat{d}_j$ (in subsequent sections, we will discuss numerical simulations that include the cubic and quartic contributions).  The problem then reduces to
\begin{equation}
  i\partial_t \vect{d} = \left(\bm{H} - i\frac{\bm{\Gamma}}{2}\right)\vect{d} + \bm{U}\,\vect{d}^{\dagger} - i\sqrt{\bm{\Gamma}}\,\vect{\eta},
  \label{eq:langevin}
\end{equation}
where $\vect{d} = (\hat{d}_1, \ldots, \hat{d}_N)^T$ is the vector of fluctuation operators, $\vect{\eta} = (\hat{\eta}_1, \ldots, \hat{\eta}_N)^T$ is the vacuum input noise satisfying $\langle \hat{\eta}_j(t)\,\hat{\eta}_k^\dagger(t') \rangle = \delta_{jk}\,\delta(t - t')$, and
\begin{align}
  \bm{H} &= \mathrm{diag}(\tilde{\omega}_1,\ldots,\tilde{\omega}_N) + [J_{jk}],  \\
  \bm{\Gamma} &= \mathrm{diag}(\gamma_1,\ldots,\gamma_N), \\
  \bm{U} &= \mathrm{diag}(2U_1\alpha_1^2,\ldots,2U_N\alpha_N^2).
  \label{eq:matrices}
\end{align}
Here $\bm{H}$ is the single-particle Hamiltonian matrix (encoding renormalized detunings and couplings), $\bm{\Gamma}$ is the loss matrix, and $\bm{U}$ is the anomalous (squeezing) coupling arising from the Kerr nonlinearity interacting with the classical field.

Equation~\eqref{eq:langevin} can be solved by the method of Green's functions, as shown in Appendix~\ref{sec:greenfun}.  In terms of the retarded Green's function $\bm{G}_\omega = (\bm{H} - i\bm{\Gamma}/2 - \omega)^{-1}$, the steady-state two-time correlation functions are
\begin{align}
  \langle \vect{d}(t)\,\vect{d}^{\dagger T}\!(0) \rangle &= \mathcal{F}_{t}^{-1}\!\left[ \bm{G}_\omega\, \bm{\Gamma}\, \bm{G}_\omega^{\dagger} \right], \label{eq:corr1} \\
  \langle \vect{d}^{\dagger}(0)\,\vect{d}^{T}\!(t) \rangle &= \mathcal{F}_{t}^{-1}\!\left[ \bm{G}_\omega^{\dagger} \bm{U}^{\dagger} \bm{G}_{-\omega}\, \bm{\Gamma}\, \bm{G}_{-\omega}^{\dagger} \bm{U}\, \bm{G}_\omega \right], \label{eq:corr2} \\
  \langle \vect{d}(t)\,\vect{d}^{T}\!(0) \rangle &= -\mathcal{F}_{t}^{-1}\!\left[ \bm{G}_\omega\, \bm{\Gamma}\, \bm{G}_\omega^{\dagger} \bm{U}\, \bm{G}_{-\omega} \right], \label{eq:corr3} \\
  \langle \vect{d}^{\dagger}(0)\,\vect{d}^{\dagger T}\!(t) \rangle &= -\mathcal{F}_{t}^{-1}\!\left[ \bm{G}_\omega^{\dagger} \bm{U}^{\dagger} \bm{G}_{-\omega}\, \bm{\Gamma}\, \bm{G}_{-\omega}^{\dagger} \right], \label{eq:corr4}
\end{align}
where $\mathcal{F}_{t}^{-1}[f(\omega)] = \int \frac{d\omega}{2\pi}\, e^{-i\omega t}\, f(\omega)$ is the usual inverse Fourier transform.

\section{Superbunching point}

We are interested in the $g^{(2)}$ function given in Eq.~\eqref{g2def}.  To simplify the four-operator correlation function in the numerator, we use Wick's theorem (which is valid for Gaussian states in the linearized theory).  At $\tau = 0$,
\begin{align*}
  \big\langle \hat{d}_i^\dagger \, \hat{d}_i^\dagger\, \hat{d}_i\, \hat{d}_i \big\rangle 
  = \big|\big\langle \hat{d}_i\, \hat{d}_i \big\rangle\big|^2
  + \big|\big\langle \hat{d}_i^\dagger\, \hat{d}_i \big\rangle\big|^2
  + \big\langle \hat{d}_i^\dagger\, \hat{d}_i \big\rangle\,
  \big\langle \hat{d}_i^\dagger\, \hat{d}_i \big\rangle,
\end{align*}
and so
\begin{equation}
  g_{ii}^{(2)}(0) = 2 + \left|\frac{\langle \hat{d}_i\, \hat{d}_i \rangle}{\langle \hat{d}_i^\dagger \hat{d}_i \rangle}\right|^2.
  \label{eq:S_g2_opt}
\end{equation}
In the second term, the numerator is given by Eq.~\eqref{eq:corr3}:
\begin{equation}
  \langle \hat{d}_i\, \hat{d}_i \rangle = -\left[\mathcal{F}_0^{-1}\!\left(\bm{G}_\omega\,\bm{\Gamma}\,\bm{G}_\omega^\dagger\,\bm{U}\,\bm{G}_{-\omega}\right)\right]_{ii},
  \label{eq:S_dd}
\end{equation}
which is of order $U\alpha^2$.

As for the numerator, let us consider the special case where the system is tuned to a point where the classical amplitude vanishes at a site $i$: $\alpha_i = 0$.  The mean photon number is then determined entirely by the zero-point fluctuations:
\begin{align}
  \langle n_i \rangle &= \langle \hat{d}_i^\dagger \hat{d}_i \rangle \\
  &= \left[\mathcal{F}_0^{-1}\!\left(\bm{G}_\omega^\dagger\,\bm{U}^\dagger\,\bm{G}_{-\omega}\,\bm{\Gamma}\,\bm{G}_{-\omega}^\dagger\,\bm{U}\,\bm{G}_\omega\right)\right]_{ii},
  \label{eq:S_n_opt}
\end{align}
where on the second step we have used Eq.~\eqref{eq:corr2}.  This is of order $U^2\alpha^4$.  Combining this with Eqs.~\eqref{eq:S_g2_opt}--\eqref{eq:corr3} gives
\begin{equation}
  g_{ii}^{(2)}(0) - 2 \;\propto\; \frac{1}{U^{2}\alpha^{4}}
  \;\propto\; \frac{1}{\langle n_i\rangle}.
  \label{eq:S_scaling}
\end{equation}

\begin{figure}
  \centering
  \includegraphics[width=\columnwidth]{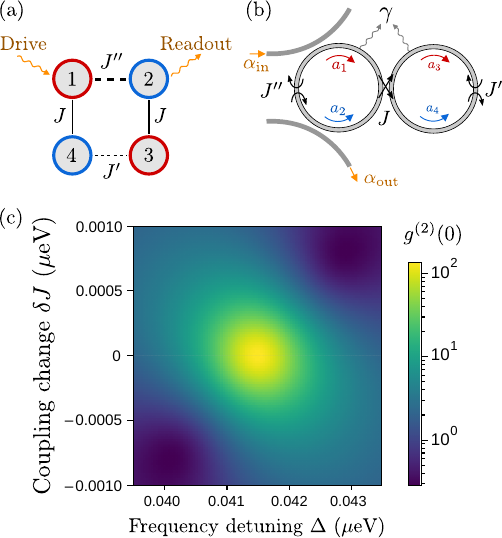}
  \caption{(a) Schematic of a four-mode model that can exhibit photon superbunching.  Each mode is subject to a weak Kerr nonlinearity; one mode $d = 1$ is coherently driven, and another mode $i = 2$ is monitored. (b) Schematic of a possible realization using two coupled ring resonators, each hosting clockwise and counterclockwise modes that are coupled by intra-ring reflection.  (c)~Numerically calculated plot of $g^{(2)}(0)$ (heat map) versus the rings' frequency detuning $\Delta$ and inter-ring coupling perturbation $\delta J$.  The fixed model parameters are $U_i=4.7\times10^{-6}\,\mu\text{eV}$, $J=J'=1\,\mu\text{eV}$, $J''=0.5\,\mu\text{eV}$, $\gamma_i=2\,\mu\text{eV}$, and $F=200\,\mu\text{eV}$.}
  \label{fig:lattice}
\end{figure}

We have thus established the phenomenon of superbunching in this weakly nonlinear system, along with an associated trade-off: it occurs around a point where the mean photon flux vanishes.  Intuitively, it is precisely when the classical field vanishes that emission is dominated by correlated photons from parametric scattering.  The practical significance of this phenomenon, therefore, depends on the photon detection limits and the maximum value of $g^{(2)}$ within those limits.

Let us now look for a configuration of optical modes that can host such a superbunching point.  Returning to Eq.~\eqref{eq:classical}, we seek a solution with $\alpha_i = 0$ (for some $i$) in the weak-nonlinearity limit $U_j|\alpha_j|^2 \ll |\Delta_j|, \gamma_j$.  By dropping the $U$ terms, we bring the equation into the linear form
\begin{align}
  \bm{L} &\begin{pmatrix}\alpha_1 \\ \vdots\\\alpha_N
  \end{pmatrix} + \begin{pmatrix}F_1\\\vdots\\F_N
  \end{pmatrix} = 0,
  \label{eq:zerocond} \\
  L_{jk} &= z_j\,\delta_{jk} + J_{jk}, \\
  z_j &\equiv -\Delta_j - i\gamma_j/2.
\end{align}
When only one mode $d$ is driven ($F_j = F\,\delta_{jd}$), the zero-displacement condition on a signal site $i \neq d$ reduces to
\begin{equation}
  L^{-1}_{id} = 0.
  \label{eq:Linvcond}
\end{equation}
By the adjugate formula $L^{-1}_{id} = \adj{\bm{L}}_{id}/\det\bm{L}$, this requires the $(i,d)$ cofactor of $\bm{L}$ to vanish.  A systematic analysis shows that for $N = 2$ or $N = 3$ modes, Eq.~\eqref{eq:Linvcond} cannot be satisfied unless at least one mode is totally lossless or has gain (see Supplemental Materials~\cite{SM}).  A simpler way to achieve such a solution is to use $N = 4$ lossy modes in a ring arrangement, as depicted in Fig.~\ref{fig:lattice}(a).  We will focus on this configuration.

The behavior away from the optimal superbunching point can also be derived within this framework.  Suppose $\alpha_i$ is not merely nonzero but dominant:
\begin{equation}
  |\alpha_j|^2 \gg \langle \hat{d}_j^\dagger \hat{d}_j \rangle.
\end{equation}
The denominator of $g^{(2)}$ then simplifies as $\langle \hat{a}_j^\dagger \hat{a}_j \rangle \approx |\alpha_j|^2$. Expanding the numerator $\langle \hat{a}_j^\dagger(0)\, \hat{a}_i^\dagger(\tau)\, \hat{a}_i(\tau)\, \hat{a}_j(0) \rangle$ and keeping only terms up to first order in $U_j\alpha_j^2$, we obtain
\begin{multline*}
  \langle \hat{a}_j^\dagger(0)\, \hat{a}_i^\dagger(\tau)\, \hat{a}_i(\tau)\, \hat{a}_j(0) \rangle \\
  \approx |\alpha_i\alpha_j|^2 - 2\operatorname{Re}\!\left\{\alpha_i^*\alpha_j^*\,\mathcal{F}_\tau^{-1}\!\left[\bm{G}_\omega\,\bm{\Gamma}\,\bm{G}_\omega^\dagger\,\bm{U}\,\bm{G}_{-\omega}\right]_{ij}\right\}.
\end{multline*}
Hence, in this regime,
\begin{equation}
  g_{ij}^{(2)}(\tau) \approx 1 - \frac{2\operatorname{Re}\!\left\{\alpha_i^*\alpha_j^*\,\mathcal{F}_\tau^{-1}\!\left[\bm{G}_\omega\,\bm{\Gamma}\,\bm{G}_\omega^\dagger\,\bm{U}\,\bm{G}_{-\omega}\right]_{ij}\right\}}{|\alpha_i\alpha_j|^2}.
  \label{eq:S_g2_away}
\end{equation}

\section{Numerical results}

We now want to verify the superbunching predictions of the previous section.  Instead of the linearized Eq.~\eqref{eq:langevin}, from which the analytic predictions were obtained, let us return to the full quantum master equation \eqref{eq:S_master}, and solve it numerically with all nonlinear terms (cubic and quartic) included.  For the model, we take the four-mode configuration of Fig.~\ref{fig:lattice}(a) and choose $d=1$ as the driven mode, and $i = 2$ as the signal mode.

We target a set of parameters applicable to the realistic setup of Fig.~\ref{fig:lattice}(b), which aims to realize the model with two coupled silicon carbide microring resonators, utilizing the clockwise and counterclockwise modes in each ring.  The setup has three sets of inter-mode couplings: a tunable inter-ring coupling $J$, and two fixed intra-ring couplings $J' = 1\,\mu\text{eV}$  and $J'' = 0.5\,\mu\text{eV}$.  All modes have the same decay rate $\gamma_j = 2\,\mu\text{eV}$ and Kerr coefficient $U_i = 4.7\times10^{-6}\,\mu\text{eV}$.  Notably, the Kerr coefficient matches the intrinsic optical nonlinearity of silicon carbide, without quantum dots or other inclusions.  The coherent drive is fixed at $F = 200\,\mu\text{eV}$, corresponding to an input power of $P_\mathrm{in} = \omega_L F / 2\pi \approx 6\,\text{nW}$ at wavelength $\lambda = 1550\,\text{nm}$.  Our chosen parameters are based on the SiC waveguide simulations; for details, see the Supplemental Materials~\cite{SM}.

With the above parameters fixed, we can vary the inter-ring coupling $J$ and detuning $\Delta$ (the same for all modes), so as to satisfy the zero-displacement condition \eqref{eq:Linvcond}.  Note that these are standard active control parameters in integrated microring platforms; in particular, fine adjustments to the effective couplings can be achieved by thermo-optic or interferometric means~\cite{deAguiar19,Yan23,Gualandi26} (see Supplemental Materials~\cite{SM}).

To perform numerical calculations of $g^{(2)}$, we first obtain the classical amplitudes $\alpha_j$ by solving Eq.~\eqref{eq:classical} using a modification of the Powell hybrid method as implemented in MINPACK~\cite{More80}. The effective Hamiltonian $\mathcal{H}_\mathrm{eff}$ and the Lindblad collapse operators are then constructed in a truncated Fock basis with per-mode photon-number cutoffs $N_1=7$, $N_2=4$, $N_3=5$, and $N_4=5$, which are later self-consistently validated to be significantly larger than the obtained steady-state per-mode photon numbers.  Setting $d\hat{\rho}/dt=0$ yields a sparse linear system on the vectorized density matrix (the Liouvillian superoperator), which is solved using the direct LGMRES iterative method implemented in QuTiP~\cite{Johansson13}.  We then evaluate $g^{(2)}(\tau)$ via the quantum regression theorem~\cite{Johansson13, Lax63}.  As a final check, we verify that the results are insensitive to further increases in the Fock cutoffs $N_j$.

It is worth noting that the aforementioned Fock space cutoffs refer to the displaced operators $\hat{d}_j$, not the original $\hat{a}_j$.  Hence, the calculations are valid and tractable even away from the zero-displacement point, where the full photon occupation $\langle \hat{n}_j \rangle$ may not be small.

Fig.~\ref{fig:lattice}(c) shows the resulting values of $g^{(2)}(0)$ as a function of the detuning $\Delta$ and coupling perturbation $\delta J$.  A single sharp peak of superbunching is found at the point where $\alpha_i = 0$.  Its maximum reaches $g^{(2)}(0) \approx 135$, which is well within the superbunched regime.  At this point, the signal cavity has mean photon occupation $\langle n_i \rangle \approx 6.7\times10^{-4}$, so for a realistic output-waveguide coupling rate of $\gamma_\mathrm{out} \approx 0.5\,\mu\text{eV}$ (see Supplemental Materials~\cite{SM}), the photon emission rate is $R_\mathrm{em} = \langle n_i \rangle \gamma_\mathrm{out} / (2\pi\hbar) \approx 80\,\text{kHz}$, compatible with commercial single-photon detectors. 

\begin{figure}
  \centering
  \includegraphics[width=\columnwidth]{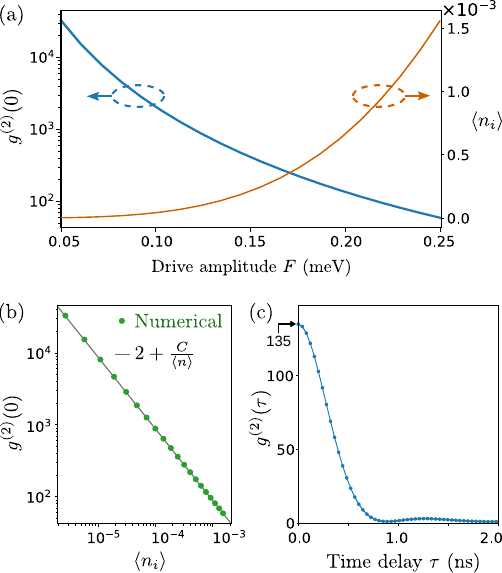}
  \caption{Characterization of superbunching solutions. (a)~Plot of $g^{(2)}(0)$ and mean photon number $\langle n\rangle$ at the optimal superbunching point for each drive amplitude $F$.  These results are obtained by solving the quantum master equation numerically.  (b) Similar to (a), but plotting $\langle n\rangle$ versus $g^{(2)}(0)$ at each optimal superbunching point (dots), along with the predicted scaling relation $g^{(2)}(0)-2 \approx C/\langle n\rangle$ with $C$ calibrated to the numerical data point at $F=250\,\mu\text{eV}$.  (c)~Time-delayed second-order correlation $g^{(2)}(\tau)$ at an optimal superbunching point with $g^{(2)}(0)\approx 135$.  In all subplots, we fix $\Delta=0.0415\,\mu\text{eV}$ and $\gamma=2\,\mu\text{eV}$.}
  \label{fig:optimum}
\end{figure}

As we have seen, there is a trade-off between the degree of superbunching that can be achieved using this method and the total photon population, since the superbunching point coincides with a zero of the classical field.  If one can tolerate lower photon emission rates (e.g., by having ultra-low detector noise levels), the trade-off may be further exploited to achieve extraordinarily large $g^{(2)}(0)$.

In Fig.~\ref{fig:optimum}(a), we plot both $g^{(2)}(0)$ and $\langle \hat{n}_i \rangle$ at the optimal superbunching point against the drive amplitude $F$ (i.e., for each $F$, the couplings and detuning are re-tuned to the optimal superbunching point).  The results indicate that if one is able and willing to operate at very low photon count rates, it is possible (say) $g^{(2)}(0) > 3\times10^4$ at $F=50\,\mu\text{eV}$, a level of superbunching previously reported only in a macroscopic nonlinear photonic crystal fiber~\cite{Qin24}.  In principle, even larger values are accessible at weaker drives.

To verify that the numerically-observed trade-off follows the theory quantitatively, Fig.~\ref{fig:optimum}(b) directly plots $g^{(2)}(0)$ against $\langle \hat{n}_i \rangle$, using the numerically-obtained values at each optimal superbunching point under varying $F$, alongside the theoretical scaling relation $g^{(2)}(0)-2 = C/\langle \hat{n}_i \rangle$ from Eq.~\eqref{eq:S_scaling}.  Here, the coefficient $C$ is chosen so that the value at $F=250\,\mu\text{eV}$ agrees with the numerical result.  Evidently, the scaling relation accurately predicts the numerical results, including in the giant superbunching regime $g^{(2)}(0) \gg 10^2$.

These predictions may be directly compared to previous experimental superbunching results on other sytems.  For optically-driven giant superbunching from a single perovskite quantum dot at cryogenic temperatures, $g^{(2)}(0) \approx 9$ has been obtained with an emission rate of $R_{\textrm{em}} \approx 533\,\text{kHz}$ (at $100\,\text{nW}$ pump power), with a maximum of $g^{(2)}(0) \approx 30$ at weaker pump powers~\cite{Wang21opt}.  For four-wave mixing in a nonlinear photonic crystal fiber, $g^{(2)}(0) \approx 1200$ has been reported with a count rate of $200\,\text{kHz}$~\cite{Petrov19}.  For cathodoluminescence from nanodiamond nitrogen-vacancy centers, $g^{(2)}(0) \approx 49$ was achieved with $R_{\textrm{em}} \sim 10\,\text{kHz}$~\cite{Feldman18}.

Therefore, our proposed approach supports two complementary operating regimes.  First, at $F=200\,\mu\text{eV}$, it can achieve larger $g^{(2)}(0)$ than several previous compact sources while maintaining a comparable emission rate of $R_{\textrm{em}} \approx 80\,\text{kHz}$.  Second, if the drive is reduced to $F=50\,\mu\text{eV}$, one can reach $g^{(2)}(0)>3\times10^4$, well above established nonlinear photonic crystal fiber results \cite{Petrov19} (and comparable only to a very recent claim based on another photonic crystal fiber \cite{Qin24})---albeit with a much lower count rate.

Finally, Fig.~\ref{fig:optimum}(c) shows the time-delayed second-order correlation $g^{(2)}(\tau)$ at the optimal superbunching point in Fig.~\ref{fig:lattice}(c). The correlation exhibits a sharp peak at $\tau = 0$ and decays on a timescale set by the cavity linewidth $\gamma$, confirming that the superbunching is a genuine quantum correlation effect rather than arising from classical intensity fluctuations.

\section{Conclusions}

We have shown that photon superbunching can be achieved in a system of coupled optical modes subject to only weak Kerr nonlinearities.  The essential physics is captured by Eq.~\eqref{eq:scaling_result}, which predicts a divergence in the second-order photon correlation as the nonlinearity strength $U$ and the photon population $\langle \hat{n}_i\rangle$ both approach zero.  It is striking that a \textit{weaker} nonlinearity induces a \textit{stronger} quantum effect, which runs counter to the conventional wisdom that strong photon correlations demand strong interactions.  The reason is that interference is utilized to suppress the classical amplitude, so as to allow quantum fluctuations to dominate.

We have presented an exemplary setup for realizing this phenomenon, based on coupled silicon carbide ring resonators with nanowatt-level driving, standard forms of active parameter tuning, and other design elements that are commonplace in integrated photonics~\cite{Yi22,Xing19}.  When the system is driven at $F = 200\,\mu\text{eV}$, the photons in the signal cavity reach $g^{(2)}(0) \approx 135$ with an emission rate $80\,\text{kHz}$, which is measurable with commercially available single-photon detectors without employing any unusual ultralow-noise or ultrahigh-efficiency techniques.  If the drive may be reduced to $F = 50\,\mu\text{eV}$, we predict $g^{(2)}(0) \gtrsim 3\times10^4$ [Fig.~\ref{fig:optimum}(a)]; higher values of $g^{(2)}(0)$ are accessible so long as one can tolerate a lower emission rate from such a light source.

So far as we know, this route to achieving arbitrary strong levels of superbunching has not previously been explored.  Although our predictions were derived from a linearized theory, they are confirmed by quantum master equation simulations retaining all nonlinear terms.

Several directions merit further investigation. Engineering cavity couplings in larger lattices~\cite{Carusotto13} may enable multiphoton superbunching through higher-order $g^{(N)}$, which is relevant to quantum-enhanced metrology and ghost imaging beyond the two-photon level. Unlike conventional schemes that rely on intrinsic emitter properties, our mechanism requires only coupled nonlinear cavities and a coherent drive; however, combining both approaches, by embedding quantum emitters in coupled cavity networks, may allow collective emission dynamics to be further engineered by intercavity interference.  This could be used to access correlation regimes inaccessible to either mechanism alone.

\appendix

\section{Effective fluctuation Hamiltonian}
\label{sec:supp_Heff}

In this Appendix, we derive the effective Hamiltonian \eqref{eq:Heff}, which serves as the basis of the superbunching prediction.  We start from the system Hamiltonian \eqref{eq:H}, which describes $N$ coherently-driven coupled modes with Kerr nonlinearities, in a co-rotating frame; the quantum fluctuation operators $\hat{d}_j$ are split out via Eq.~\eqref{eq:displace}, with the mean fields given by Eq.~\eqref{eq:classical}.  To proceed, we expand the Hamiltonian, substituting Eq.~\eqref{eq:displace} into each term of the Hamiltonian and collect contributions by order in $\hat{d}_j$.

The first (detuning) term in the Hamiltonian yields
\begin{align}
  -\Delta_j \hat{a}_j^\dagger \hat{a}_j
  &= -\Delta_j \left[|\alpha_j|^2 + \alpha_j^* \hat{d}_j + \alpha_j \hat{d}_j^\dagger + \hat{d}_j^\dagger \hat{d}_j\right].
\end{align}
For the second term, which corresponds to the Kerr nonlinearity, we expand $(\alpha_j^* + \hat{d}_j^\dagger)^2(\alpha_j + \hat{d}_j)^2$ exactly:
\begin{align}
  &U_j \hat{a}_j^\dagger \hat{a}_j^\dagger \hat{a}_j \hat{a}_j \nonumber \\
  &= U_j \bigg[|\alpha_j|^4 + 2|\alpha_j|^2 \alpha_j^* \hat{d}_j + 2|\alpha_j|^2 \alpha_j \hat{d}_j^\dagger \nonumber \\
  &\quad\qquad + 4|\alpha_j|^2 \hat{d}_j^\dagger \hat{d}_j + \alpha_j^2 (\hat{d}_j^\dagger)^2 + (\alpha_j^*)^2 \hat{d}_j^2 \nonumber \\
  &\quad\qquad + 2\alpha_j^* \hat{d}_j^\dagger \hat{d}_j^2 + 2\alpha_j (\hat{d}_j^\dagger)^2 \hat{d}_j + (\hat{d}_j^\dagger)^2 \hat{d}_j^2\bigg].
  \label{eq:S_kerr_expand}
\end{align}
For the coupling terms, we obtain
\begin{align}
  J_{jk} \hat{a}_j^\dagger \hat{a}_k
  &= J_{jk} \left[\alpha_j^* \alpha_k + \alpha_j^* \hat{d}_k + \alpha_k \hat{d}_j^\dagger + \hat{d}_j^\dagger \hat{d}_k\right].
\end{align}
Finally, the driving terms give
\begin{align}
  F_j \hat{a}_j^\dagger + F_j^* \hat{a}_j
  &= F_j \alpha_j^* + F_j^* \alpha_j + F_j \hat{d}_j^\dagger + F_j^* \hat{d}_j.
\end{align}

We now collect terms by order.  The zeroth order terms correspond to a constant energy shift, which only affects the global phase and therefore may be ignored.
As for the first order terms, we find that the coefficients exactly match the classical steady-state equation~\eqref{eq:classical}, so as to form
\begin{align}
  \mathcal{H}_1 = \sum_j \frac{i\gamma_j}{2}(\alpha_j \hat{d}_j^\dagger - \alpha_j^* \hat{d}_j) + \mathrm{h.c.},
\end{align}
where ``$\textrm{h.c.}$'' stands for ``Hermitian conjugate''.  We should also note that the displacement transformation modifies the Lindblad dissipator:
\begin{equation}
  \gamma_j\mathcal{D}[\alpha_j + \hat{d}_j]\rho = \gamma_j\mathcal{D}[\hat{d}_j]\rho - i[\mathcal{H}_\mathrm{diss}^{(j)}, \rho],
\end{equation}
where
\begin{equation}
  \mathcal{H}_\mathrm{diss}^{(j)} = \frac{i\gamma_j}{2}\!\left(\alpha_j^* \hat{d}_j - \alpha_j \hat{d}_j^\dagger\right).
\end{equation}
Hence, the linear terms cancel exactly between the Hamiltonian and dissipative contributions:
\begin{equation}
  \mathcal{H}_1 + \sum_j \mathcal{H}_\mathrm{diss}^{(j)} = 0,
\end{equation}
which self-consistently satisfies $\langle \hat{d}_j \rangle = 0$.

Thus, the fluctuation dynamics starts at quadratic order.  Collecting these terms gives
\begin{equation}
  \mathcal{H}_2 = \sum_j \left[\tilde{\omega}_j\, \hat{d}_j^\dagger \hat{d}_j + U_j\!\left(\alpha_j^2 (\hat{d}_j^\dagger)^2 + \textrm{h.c.}\right)\right] + \sum_{j \neq k} J_{jk}\, \hat{d}_j^\dagger \hat{d}_k,
\end{equation}
where $\tilde{\omega}_j$ is the renormalized frequency detuning given in Eq.~\eqref{eq:omegatilde}.  The terms of order $\alpha^2$ describe squeezing (parametric) processes induced by the classical field.

Next, we handle the the third- and fourth-order terms, which arise solely from the Kerr nonlinearity:
\begin{align}
  \mathcal{H}_3 &= \sum_j 2U_j \left(\alpha_j^*\, \hat{d}_j^\dagger \hat{d}_j^2 + \alpha_j\, (\hat{d}_j^\dagger)^2 \hat{d}_j\right).
  \label{eq:S_H3} \\
  \mathcal{H}_4 &= \sum_j U_j\, \hat{d}_j^\dagger \hat{d}_j^\dagger \hat{d}_j \hat{d}_j.
  \label{eq:S_H4}
\end{align}
Combining all non-constant orders, we obtain the effective Hamiltonian \eqref{eq:Heff}.  Note that this result is exact---we have not dropped any terms in the expansion.

\section{The linearized Langevin equation}
\label{sec:greenfun}

Here, we give the derivation leading to Eqs.~\eqref{eq:corr1}--\eqref{eq:corr4}.  Starting from the effective Hamiltonian \eqref{eq:Heff} and the master equation \eqref{eq:S_master}, we retain only terms up to quadratic order in $\hat{d}_j$.  The quantum Langevin equations become
\begin{align}
  i\partial_t \hat{d}_k &= \!\left(\tilde{\omega}_k - i\frac{\gamma_k}{2}\right)\! \hat{d}_k + 2U_k \alpha_k^2 \hat{d}_k^\dagger
  + \sum_l J_{kl} \hat{d}_l - i\gamma_k^{\frac{1}{2}}\hat{\eta}_k,
  \label{eq:S_langevin_single}
\end{align}
where $\hat{\eta}_k$ is the vacuum input noise operator satisfying $\langle \hat{\eta}_j(t)\, \hat{\eta}_k^\dagger(t') \rangle = \delta_{jk}\,\delta(t-t')$ and $\langle \hat{\eta}_j(t)\, \hat{\eta}_k(t') \rangle = 0$.  In Eq.~\eqref{eq:langevin}--\eqref{eq:matrices}, these equations were casted into matrix form by introducing $\vect{d}$ (collecting the $\hat{d}_j$ operators), as well as the matrices $\bm{H}$, $\bm{\Gamma}$, and $\bm{U}$ (collecting the detunings/couplings, loss rates, and nonlinearity).  Similarly, taking the Hermitian conjugate of Eq.~\eqref{eq:S_langevin_single} yields
\begin{equation}
  -i\partial_t \vect{d}^\dagger = (\bm{H} + i\bm{\Gamma}/2)\,\vect{d}^\dagger + \bm{U}^*\,\vect{d} + i\sqrt{\bm{\Gamma}}\,\vect{\eta}^\dagger.
  \label{eq:S_langevin_adj}
\end{equation}
Eqs.~\eqref{eq:langevin} and \eqref{eq:S_langevin_adj} combine into a Bogoliubov equation
\begin{multline}
  i\partial_t \begin{pmatrix} \vect{d} \\ \vect{d}^\dagger \end{pmatrix} =
  \begin{pmatrix} \bm{H} - i\bm{\Gamma}/2 & \bm{U} \\ -\bm{U}^* & -(\bm{H} + i\bm{\Gamma}/2) \end{pmatrix}
  \begin{pmatrix} \vect{d} \\ \vect{d}^\dagger \end{pmatrix}
  \\ - i \begin{pmatrix} \sqrt{\bm{\Gamma}}\vect{\eta} \\ \sqrt{\bm{\Gamma}}\vect{\eta}^\dagger \end{pmatrix}.
  \label{eq:S_2N}
\end{multline}
To solve this, we define the Fourier transformed operators
\begin{equation}
  \hat{d}_k(\omega) = \int dt\, e^{i\omega t}\, \hat{d}_k(t).
\end{equation}
In the frequency domain, Eq.~\eqref{eq:S_2N} becomes
\begin{equation}
  \begin{pmatrix} -i\vect{d}_\omega \\ i\vect{d}_{-\omega}^\dagger \end{pmatrix} =
  \bm{\mathcal{M}}_\omega^{-1}
  \begin{pmatrix} \sqrt{\bm{\Gamma}}\,\vect{\eta}_\omega \\ \sqrt{\bm{\Gamma}}\,\vect{\eta}_{-\omega}^\dagger \end{pmatrix},
  \label{eq:S_fourier_sol}
\end{equation}
where
\begin{align}
  \bm{\mathcal{M}}_\omega &= \begin{pmatrix} \bm{H} - i\bm{\Gamma}/2 - \omega & -\bm{U} \\ -\bm{U}^* & \bm{H} + i\bm{\Gamma}/2 + \omega \end{pmatrix},
  \label{eq:S_Mmatrix} \\
  \langle \vect{\eta}_{\omega_1} \vect{\eta}_{-\omega_2}^{\dagger T} \rangle &= 2\pi\,\delta(\omega_1 + \omega_2)\,\mathbb{I}, \quad
  \langle \vect{\eta}_{\omega_1} \vect{\eta}_{\omega_2}^T \rangle = 0.
  \label{eq:S_noise_corr}
\end{align}
Next, let us define the retarded Green's function as the inverse of the upper-left block in $\bm{\mathcal{M}}_\omega$:
\begin{equation}
  \bm{G}_\omega = (\bm{H} - i\bm{\Gamma}/2 - \omega)^{-1}.
  \label{eq:S_Green}
\end{equation}
Note that for real $\omega$,
\begin{align}
  \bm{G}_{-\omega}^\dagger = (\bm{H} + i\bm{\Gamma}/2 + \omega)^{-1}.
\end{align}
We can then compute the inverse of $\bm{\mathcal{M}}_\omega$ via the Schur complement.
To leading order in the nonlinearity term $\bm{U}$, Eq.~\eqref{eq:S_fourier_sol} reduces to
\begin{align}
  -i\vect{d}_\omega &\approx \bm{G}_\omega \sqrt{\bm{\Gamma}}\,\vect{\eta}_\omega + \bm{G}_\omega\, \bm{U}\, \bm{G}_{-\omega}^\dagger \sqrt{\bm{\Gamma}}\,\vect{\eta}_{-\omega}^\dagger, \label{eq:S_dsol}\\
  i\vect{d}_{-\omega}^\dagger &\approx \bm{G}_{-\omega}^\dagger\, \bm{U}^\dagger\, \bm{G}_\omega\, \sqrt{\bm{\Gamma}}\,\vect{\eta}_\omega + \bm{G}_{-\omega}^\dagger\sqrt{\bm{\Gamma}}\,\vect{\eta}_{-\omega}^\dagger. \label{eq:S_ddag_sol}
\end{align}
Using Eq.~\eqref{eq:S_noise_corr}, we can calculate correlators like
\begin{align*}
  \langle \vect{d}_{\omega_1}\, \vect{d}_{-\omega_2}^T \rangle
  &= -2\pi\, \bm{G}_{\omega_1} \bm{\Gamma}
  \big[\bm{G}_{\omega_2}\, \bm{U}\, \bm{G}_{-\omega_2}^\dagger\big]^T
  \;\delta(\omega_1+\omega_2),
\end{align*}
and, likewise, $\langle \vect{d}_{-\omega_1}^\dagger\, \vect{d}_{\omega_2}^T \rangle$, $\langle \vect{d}_{-\omega_1}^\dagger\, \vect{d}_{-\omega_2}^{\dagger T} \rangle$, and $\langle \vect{d}_{\omega_1}\, \vect{d}_{-\omega_2}^{\dagger T} \rangle$.   Finally, Fourier transforming back to the time domain yields the correlation functions \eqref{eq:corr1}--\eqref{eq:corr4}.

\bibliography{ref}

\clearpage

\begin{widetext}

\makeatletter 
\renewcommand{\theequation}{S\arabic{equation}}
\makeatother
\setcounter{equation}{0}

\makeatletter 
\renewcommand{\thefigure}{S\@arabic\c@figure}
\makeatother
\setcounter{figure}{0}

\makeatletter 
\renewcommand{\thesection}{S\arabic{section}}
\makeatother
\setcounter{section}{0}

\setcounter{page}{1}

\begin{center}
  {\large Supplemental Materials for} \\ \vskip 0.1in
  {\Large Giant Photon Superbunching from Weak Nonlinearity} \\
  \vskip 0.1in
  Y.~Wang, X.~Zheng, T.~C.~H.~Liew, and Y.~D.~Chong
\end{center}
  
\section{Minimum number of cavities for zero displacement}
\label{sec:supp_mincav}

Here we show that a minimum of four cavities is required to satisfy the zero-displacement condition $L^{-1}_{id} = 0$ [Eq.~\eqref{eq:Linvcond} of the main text] under the practical requirement that every cavity has nonzero loss ($\gamma_j > 0$) and the inter-cavity couplings $J_{jk}$ are real and nonvanishing.  We consider a general scenario in which all couplings and the complex detunings $z_j = -\Delta_j - i\gamma_j/2$ are independently tunable.

\subsection{Two cavities}

For two cavities coupled by $J$:
\begin{equation}
  \bm{L} = \begin{pmatrix} z_1 & J \\ J & z_2 \end{pmatrix}, \quad
  \bm{L}^{-1} = \frac{1}{z_1 z_2 - J^2} \begin{pmatrix} z_2 & -J \\ -J & z_1 \end{pmatrix}.
  \label{eq:S_L2}
\end{equation}
The off-diagonal element $L^{-1}_{12} = -J/(z_1 z_2 - J^2)$ vanishes only if $J = 0$ (trivially decoupled).  The diagonal element $L^{-1}_{11} = z_2/(z_1 z_2 - J^2)$ vanishes only if $z_2 = 0$, i.e., $\Delta_2 = 0$ and $\gamma_2 = 0$, requiring a lossless on-resonance cavity.  Neither option is practical.

\subsection{Three cavities}

For three cavities with all pairwise couplings $J_1$, $J_2$, $J_3$:
\begin{equation}
  \bm{L} = \begin{pmatrix} z_1 & J_1 & J_3 \\ J_1 & z_2 & J_2 \\ J_3 & J_2 & z_3 \end{pmatrix}.
  \label{eq:S_L3}
\end{equation}
The cofactors entering $\bm{L}^{-1}$ are $L^{-1}_{ij} = \adj{\bm{L}}_{ij}/\det\bm{L}$, where
\begin{align}
  \adj{\bm{L}}_{11} &= z_2 z_3 - J_2^2, \label{eq:S_adj3_diag} \\
  \adj{\bm{L}}_{12} &= J_2 J_3 - z_3 J_1, \label{eq:S_adj3_off}
\end{align}
and the remaining cofactors follow by permutation.  There are two cases:

(i)~\textit{Diagonal zero.}  Setting $\adj{\bm{L}}_{11} = 0$ requires $z_2 z_3 = J_2^2$.  Taking modulus and argument separately: $|z_2|\,|z_3| = J_2^2$ and $\arg(z_2) + \arg(z_3) = 0$.  Since $\mathrm{Im}(z_j) = -\gamma_j/2 < 0$, the argument condition forces $z_2 = z_3^*$, which demands that one cavity has gain precisely balanced against another's loss, or that both cavities are lossless.

(ii)~\textit{Off-diagonal zero.}  Setting $\adj{\bm{L}}_{12} = 0$ requires $z_3 = J_2 J_3/J_1$, which is purely real.  This is only possible when $\gamma_3 = 0$---the cavity must be entirely lossless.

Both requirements are impractical in typical photonic systems.

\subsection{Four cavities in a ring}

We now consider four cavities arranged in a ring with nearest-neighbor couplings $J_{12}$, $J_{23}$, $J_{34}$, $J_{41}$ and no cross-couplings:
\begin{equation}
  \bm{L} = \begin{pmatrix} z_1 & J_{12} & 0 & J_{41} \\ J_{12} & z_2 & J_{23} & 0 \\ 0 & J_{23} & z_3 & J_{34} \\ J_{41} & 0 & J_{34} & z_4 \end{pmatrix}.
  \label{eq:S_L4}
\end{equation}
Taking cavity~1 as the driven site and cavity~2 as the signal site, the zero-displacement condition is $L^{-1}_{21} = 0$, i.e., $\adj{\bm{L}}_{21} = 0$.  Evaluating the $(2,1)$ cofactor:
\begin{equation}
  \adj{\bm{L}}_{21} = -\bigl[J_{12}(z_3 z_4 - J_{34}^2) + J_{41} J_{34} J_{23}\bigr].
  \label{eq:S_adj21_4cav}
\end{equation}
Setting this to zero gives
\begin{equation}
  z_3 z_4 = J_{34}^2 - \frac{J_{41}\, J_{34}\, J_{23}}{J_{12}}.
  \label{eq:S_4cav_cond}
\end{equation}
The right-hand side is real, while the left-hand side $z_3 z_4 = (-\Delta_3 - i\gamma_3/2)(-\Delta_4 - i\gamma_4/2)$ has imaginary part $-(\Delta_3\gamma_4 + \Delta_4\gamma_3)/2$, which can be zeroed by choosing appropriate detunings.  For identical cavities ($z_j = z = -\Delta - i\gamma/2$), the condition becomes $z^2 = J_{34}^2 - J_{41} J_{34} J_{23}/J_{12}$, whose imaginary part $\Delta\gamma = 0$ vanishes at $\Delta = 0$, leaving the purely real condition
\begin{equation}
  \frac{\gamma^2}{4} = \frac{J_{41}\, J_{34}\, J_{23}}{J_{12}} - J_{34}^2.
  \label{eq:S_gamma_cond}
\end{equation}
This is readily satisfied when $J_{41} J_{23}/J_{12} > J_{34}$, demonstrating that the four-cavity ring admits zero-displacement solutions with all cavities having nonzero loss.  Four is therefore the minimum number of cavities that enables giant superbunching under practical conditions.

\begin{figure}
  \centering
  \includegraphics[width=0.9\columnwidth]{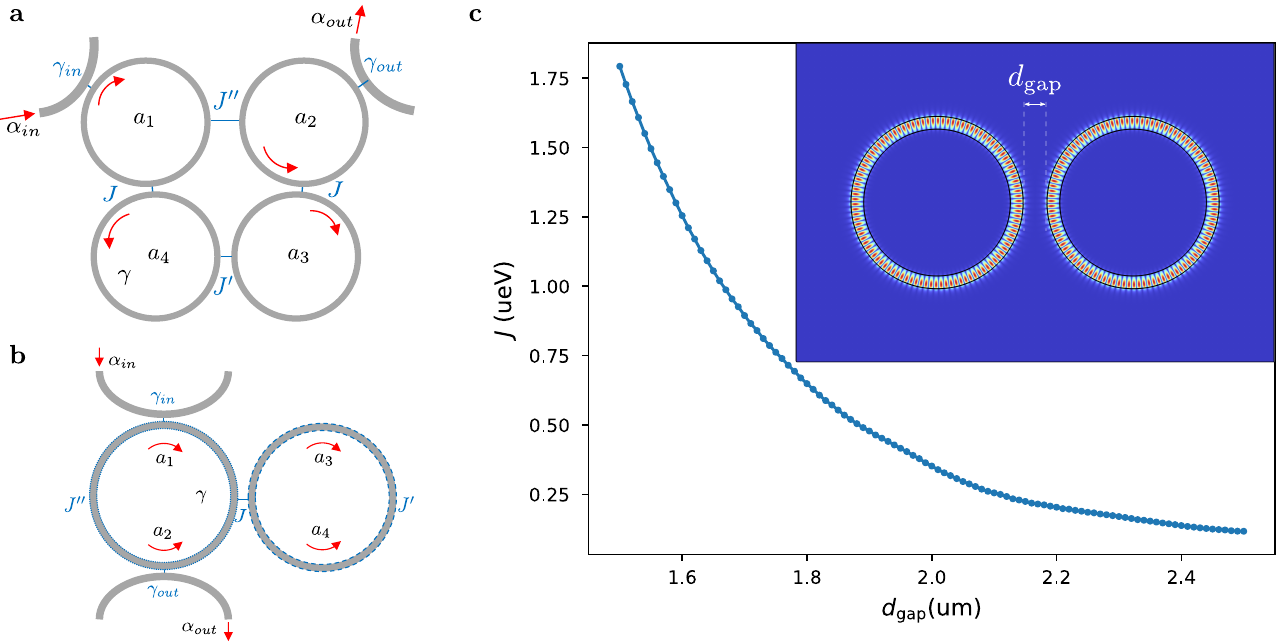}
  \caption{Ring-resonator realizations of the four-cavity model. (a)~Realization using four ring resonators. (b)~Realization using two ring resonators, with their CW and CCW modes providing the four modes. (c)~FEM simulation of the inter-ring coupling $J$ versus gap distance $d_\text{gap}$. Inset: FEM simulation of the mode profile of two coupled ring resonators at gap distance $d_\text{gap}$.}
  \label{figS:rings_gap}
\end{figure}

\section{Physical realization using ring resonators}
\label{sec:supp_rings}

To achieve the four coupled optical modes of our model, we can employ either four coupled microring resonators, as shown in Fig.~\ref{figS:rings_gap}(a), or two coupled microring resonators in which the clockwise (CW) and counterclockwise (CCW) modes of each ring serve as the four bosonic modes, as shown in Fig.~\ref{figS:rings_gap}(b), or Fig.~\ref{fig:lattice}(b) of the main text. We focus on the two-ring realization.

Here, the CW and CCW modes of each ring have the same natural frequency, and are coupled to each other with rates $J'$ (second ring) and $J''$ (first ring). In practice, such couplings can be achieved via deliberately-introduced subwavelength defects or surface roughness~\cite{Little97,Morichetti10,Li12,Li20}. Assuming both rings have the same shape and size, the Langevin equations of motion for the four modes $\hat{a}_1$ (CW, ring 1), $\hat{a}_2$ (CCW, ring 1), $\hat{a}_3$ (CW, ring 2), $\hat{a}_4$ (CCW, ring 2) are
\begin{equation}
i\frac{d}{dt}\begin{pmatrix}\hat{a}_1\\\hat{a}_2\\\hat{a}_3\\\hat{a}_4\end{pmatrix}=
\begin{pmatrix}
z_1 & J'' & 0 & J\\
J'' & z_1 & J & 0\\
0 & J & z_2 & J'\\
J & 0 & J' & z_2
\end{pmatrix}
\begin{pmatrix}\hat{a}_1\\\hat{a}_2\\\hat{a}_3\\\hat{a}_4\end{pmatrix}
+\begin{pmatrix}i\sqrt{\gamma_{\mathrm{in}}}\,\alpha_{\mathrm{in}}\\0\\0\\0\end{pmatrix},
\label{eq:S_langevin_ring}
\end{equation}
where $z_1 = -\Delta - i(\gamma+\gamma_{\mathrm{in}}+\gamma_{\mathrm{out}})/2$, $z_2 = -\Delta - i\gamma'/2$, $\Delta=\omega_c-\omega_L$ is the cavity–laser detuning, $\gamma$ is the intrinsic loss rate, $\gamma_{\mathrm{in}}$ is the input-waveguide coupling rate, $\gamma_{\mathrm{out}}$ is the output-waveguide coupling rate, $J$ is the inter-ring coupling, and $\alpha_{\mathrm{in}}$ is the input field amplitude. The output field is $\alpha_{\mathrm{out}}=\sqrt{\gamma_{\mathrm{out}}}\,\hat{a}_2$. The Kerr nonlinear Hamiltonian for the four modes is
$\mathcal{H}_{nl} = U\sum_{i=1}^{4} \hat{a}_i^{\dagger}\hat{a}_i^{\dagger}\hat{a}_i \hat{a}_i$, where $U$ is the Kerr coefficient of each mode.

We perform Finite Element Method (FEM) simulations using material parameters drawn from the experimental literature. We assume each ring is composed of silicon carbide (SiC) with ring radius $R=3\,\mu$m, waveguide thickness $t=0.35\,\mu$m, and waveguide width $w=0.8\,\mu$m, operating at wavelength $\lambda\approx1550\,\text{nm}$. The refractive indices are $n_\mathrm{SiC}=2.45$ for the waveguides and $n(\text{SiO}_2)=1.44$ for the cladding, and the nonlinear refractive index for SiC is $n_2=4.8\times10^{-6}\,\mu\text{m}^2/\text{W}$. The Kerr coefficient is estimated by~\cite{Ferretti12,Flayac16}
\begin{equation}
U=\frac{c\hbar^2\omega_c^2}{n_\mathrm{SiC}^2\,V}\,n_2,
\end{equation}
where the mode volume is approximated by the ring volume $V=2\pi Rwt$. This gives $U\approx4.7\times10^{-6}\,\mu\text{eV}$, consistent with the value used in the main text simulations.

Fig.~\ref{figS:rings_gap}(c) shows the inter-ring coupling strength $J$ as a function of gap distance $d_\mathrm{gap}$, extracted from FEM simulations. In each simulation, the eigenvalue splitting between the symmetric and antisymmetric supermodes of the coupled system equals $2J$. The FEM identifies an eigenmode near $\lambda=1553\,\text{nm}$ with a simulated intrinsic quality factor $Q_\mathrm{int}\approx1.5\times10^6$; accounting for additional loss channels including waveguide coupling ($\gamma_\mathrm{in}, \gamma_\mathrm{out}\approx 0.5\,\mu$eV), the loaded loss rate of ring-1 modes ($\hat{a}_1$, $\hat{a}_2$) is $\gamma+\gamma_\mathrm{in}+\gamma_\mathrm{out}$, while ring-2 modes ($\hat{a}_3$, $\hat{a}_4$) have only the intrinsic rate $\gamma'$ (loaded $Q\approx4\times10^5$). For simplicity, all four modes are assigned the same loss rate $\gamma_j = 2\,\mu$eV in the simulations, corresponding to the loaded quality factor of ring~1. The general theoretical model supports site-dependent $\gamma_j$ and this simplification does not qualitatively affect the conclusions. The CW\textendash CCW couplings are set to $J'=1\,\mu\text{eV}$ and $J''=0.5\,\mu\text{eV}$, which are achievable via subwavelength surface defects~\cite{Little97,Li12,Li20,Morichetti10}.

In an actual experiment, the system parameters may deviate from nominal values, so it is desirable to have $\Delta$ and $J$ actively tunable, so as to achieve the superbunching condition:

\begin{itemize}
\item The detuning $\Delta$ can be adjusted simply via the laser frequency $\omega_L$.

\item The coupling $J$ is determined principally by the gap distance $d_\mathrm{gap}$ [Fig.~\ref{figS:rings_gap}(c)], but can be further fine-tuned by engineering the refractive index in and around the gap region. In integrated microcavities, such tuning can be implemented with an auxiliary control laser~\cite{Hu20}; laser-induced heating and photoexcited carriers can modify the refractive index of semiconductors~\cite{Jensen85,Yu06}, and hence the inter-ring coupling.

In Fig.~\ref{fig:sim}(b), we illustrate the dependence of $J$ on the regional refractive index change $\Delta n$ within a spot of radius $r=0.5$ $\mu$m centered on the inter-ring gap, assuming an initial gap size of $d_{\textrm{gap}}=1.74$ $\mu$m.  For realistic levels of $\Delta n$, it is possible to achieve variations in $J$ compatible with our proposal.
\end{itemize}

\begin{figure}
    \centering
    \includegraphics[width=\textwidth]{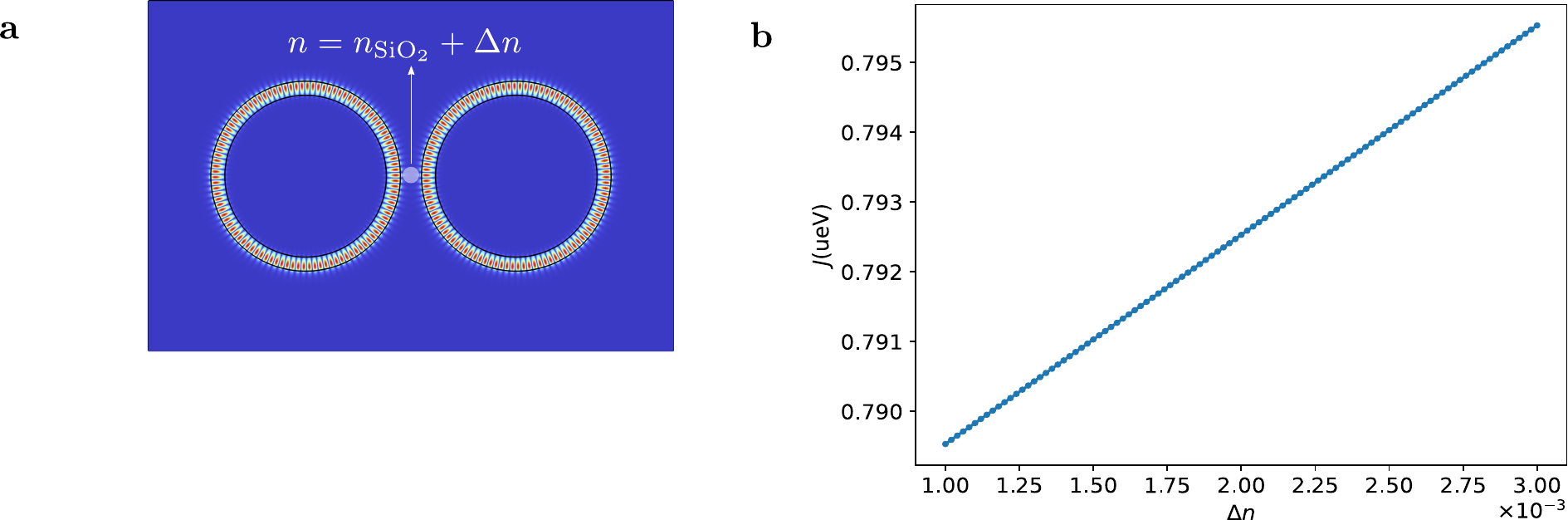}
    \caption{Fine-tuning and simulation results. (a) Schematic illustrating the $r=0.5$ $\mu$m spot within the inter-ring gap used for refractive index-based fine-tuning of the coupling $J$.  (b) Dependence of the coupling strength $J$ on the regional refractive index change $\Delta n$ applied to the spot shown in (a), assuming a fixed gap size $d_{\textrm{gap}}=1.74$ $\mu$m.}
    \label{fig:sim}
\end{figure}

\clearpage
\end{widetext}

\end{document}